\documentclass[a4paper,11pt]{article}
\usepackage{amsfonts,latexsym,amssymb}

\renewcommand{\le}{\leqslant}
\renewcommand{\ge}{\geqslant}
\renewcommand{\hat}{\widehat}
\renewcommand{\phi}{\varphi}
 \renewcommand{\P}{P_{r_N,N}}

\newcommand{\R}{\mathbf{R}} 
\newcommand{\N}{\mathbf{N}}

\newtheorem{lemma}{Lemma}
\newtheorem{theorem}{Theorem}

\newtheorem{definition}{Definition}
\newtheorem{remark}{Remark}

\def\piB{\pi_{\beta,B}}

\begin{document}
\title{\bf Strand separation in negatively supercoiled DNA}
\author{
Christian Mazza\\
LaPCS,
UFR de Math\'ematiques\\
Universit\'e Claude Bernard Lyon--1\\
B\^atiment Recherche B, Gerland\\
50 Avenue Tony Garnier\\
69366 Lyon Cedex 07, France.\\
e-mail:
christian.mazza@univ-lyon1.fr}


\maketitle

\setcounter{page}{0}

\thispagestyle{empty}

\begin{abstract}
  We consider Benham's model for strand
  separation in negatively supercoiled 
  circular DNA, and study denaturation
  as function of the superhelical
  density $\kappa <0$. 
  We propose
  a statistical version of this model, based
  on bayesian segmentation methods of current use
  in bioinformatics; this leads to new algorithms
  with priors adapted to supercoiled DNA, taking
  into account the random nature of the free energies
  needed to denature hydrogen bonds.
\end{abstract}

\vfill\vfill


\noindent
{\bf Running title:} Strand separation

\vfill

\pagebreak



\section{Introduction}
\label{s.intro}

Initiation of transcription in DNA requires the two strands
of the  double helix to separate, and strand separation
is enhanced in negative supercoiled DNA. Benham(1979,1990,1996)
proposes a mathematical model for the process of strand
separation, based on statistical mechanics ideas, and develop
algorithms to locate interesting sites along the DNA
where strand separation or replication is strongly
favored (see e.g. Clote and Backhofen(2000)). These computational methods are
Metropolis dynamics (Sun {\it et altri}(1995))
or exact methods relying on transfer matrices(Fye and Benham(1999)).
In a typical state, some hydrogen bonds are broken;
 we are interested in the repartition of the
droplets of denatured bonds,
 and on the nature of the bases situated in these
 domains.
  Our aim is
to investigate the effect of the
number of droplets, of
 the degreee of negative
superhelicity of the DNA and of the concentration
in A+T bonds on the equilibrium
properties of Benham's model. This is the topic of
Section \ref{s.denaturation}, where homopolymers
are treated according to the number of connected
domains of denatured bonds: in Section \ref{Laplace},
we study denaturation when no restriction on the
number of domains is imposed, and show that no
robust and stable denatured state exists; this is
the situation adapted to the algorithms of Fye
and Benham(1999). In Section \ref{s.s.peri},
we study the model when the number of domains
$r_N$ is such that $r_N/N\to 0$, as
$N\to\infty$, where $N$ denotes the number of bases
of the DNA. We show the existence of a stable and
robust denatured state when the duplex is sufficiently
negatively supercoiled. This is the regime
where the MCMC of Sun {\it et altri} applies. We next turn
to heteropolymers in Section \ref{Hetero}, and study
localized denaturation as function of the proportion
of A+T bonds and of the level of negative
superhelicity of the DNA. Section \ref{s.statistical}
focus on the statistical aspects of the model:
Section \ref{s.bayesian} translates Benham's model
in a Bayesian framework, and Section \ref{s.segmentation} shows
how Bayesian segmentation methods of current
use in bioinformatics, as presented in Liu and Lawrence(1999), can
be of  interest in the strand separation problem. This also
give new algorithms with priors adapted to supercoiled DNA, which take into account the random
fluctuations of the free energies needed to denature
A+T and G+C bonds.

In what
follows, we consider a spin system based
on a circular graph with node set $S$, $\vert S\vert=N$, $N\in \N$,
 and, for each
site $i\in S$,
 a spin $\sigma_i\in\{-1,+1\}$.
Benham's original model deals with a lattice gas,
with binary random variables $n_i$. We use both
notations by setting $n_i=(\sigma_i+1)/2$, and use  spins
 to link this model with known
mean field models.
The meaning of $n_i=0 $
(resp. $\sigma_i=-1$)
 is that the
bases of the double helix at site $i\in S$
are linked by an hydrogen bond,
  the link   is closed,
 and $n_i=1$
(resp. $\sigma_i=+1$) means that this bond
is broken, the link is open.
Then $n:=\sum_{i=1}^N n_i$
denotes the number of   open bonds.
 The partitioning of the DNA in  domains
 allows the linking numbers to be regulated,
 where the linking number $L$ of 
 a configuration describes the way the duplex
 winds about the axis, assuming that the axis of
 the helix is planar.
 The twist $T$ is the number of 
 times the duplex revolves about its
 axis(see e.g. Clote and
 Backhofen(2000), chap. 6.2). When the DNA is relaxed, the so-called
 B-DNA state,  a segment of $N$ bases
 produces typically
 the characteristic linking number
 $L_0=N/A$, where
 the constant $A$ is experimentally
 situated around $10.4$.
 A negative supercoilded
 DNA is a configuration
  obtained from the B-DNA by cutting
 the strands using topoisomerases of type II;
 the strands then rotate around each others
 in the direction opposite of the twist of
 the helix, reducing then its linking number
 ($L<L_0$).
 This forces then the circular axis of the helix
 to wind, producing then a more twisted and
 compact configuration(see e.g. Lewin(1994)).
This supercoiled state
 permits for example to put the helix in
 nuclei.  Benham's model permits to quantify
 the way supercoiling enhances strand separation.

 Consider a supercoiled DNA with negative
 linking difference $\alpha=L-L_0<0$, imposed during
 the process. Assume that $n$ links between bases
 are broken; the helix unwinds locally and thus increases
 its linking number to $\alpha+n/A$. Because the strands
 car rotate around each others, the same process
 induced also a twist ${\cal T}$, yielding a residual
 linking difference $\alpha_r=\alpha+n/A-{\cal T}$.
 The total twist ${\cal T}$ between separated regions
 is modeled as
 $${\cal T}=\sum_{i=1}^N \frac{n_i \tau_j}{2\pi},$$
 where $\tau_i\in\R$ is the local helicity.
 Benham's idea is to quantify all of these steps with
 free energy costs.  The torsional free energy $G_t$
 is given by
 $$G_t= \frac{C}{2}\sum_{i=1}^N n_i \tau_i^2,$$
 for some stiffness coefficient $C>0$.
 Consider
 a configuration in which $n$ bonds are separated in
 $r$ runs, that is in $r$ connected components of
 open bonds. In what follows, $2r$ is the
 {\it perimeter of the configuration}.
  
  The free energy cost for separation is modeled
 as
 $$G_s=a r+\sum_{i=1}^N b_i n_i,\ a>0,\ r=\sum_i n_i (1-n_{i+1}),$$
 where the parameters $b_i$ indicate
 the natures of the bases located at sites $i\in S$:
 $AT$ links
  are formed of 2 hydrogen bonds, and
 GC  links consist in 3 hydrogen bonds.
 If $i\in S$ is associated with an AT link, we set
 $b_i=b_{AT}$ and
 $b_i=b_{GC}$ otherwise, with
$b_{GC}>b_{AT}$. In the homopolymer case,
$b_i\equiv b>0$. This will act as an exterior
magnetic field in the statistical approach; when the
bases are chosen at random on the DNA with
some law,  $\sum_i n_i b_i$ can be viewed as a
random external field.
The free energy $b$ needed to break the 
hydrogen bonds, and thus to separate a base pair,
depends on the inverse temperature $\beta$:
\begin{equation}\label{FreeEnergy}
b=b(\beta)=\triangle H(1-\frac{\beta_m}{\beta} ),
\end{equation}
where $\triangle H$ is the enthalpy of the reaction
and $\beta_m$ is the inverse temperature associated
to the melting temperature $T_m$. Below $T_m$, the
field $b$ is positive, this is the regime we are
interested in. From Benham(1992), the pBR322 DNA
is such that $\triangle H_{AT}=7.25$ kcal/mol, 
$\triangle H_{GC}=9.02 $ kcal/mol, and the melting
temperature $T_m$ follows the law
$$T_m=354.55 +16.6 \log(x)+41 F_{GC},$$
where $x$ is some parameter and
$F_{GC}=0$ for AT bonds and $F_{GC}=1$
for GC bonds. When $x=0.01$ and $T=\beta^{-1}=310 $ K,
the resulting free energies are given by
$b_{AT}=0.255$ kcal/mol and $b_{GC}=1.301$ 
kcal/mol.

Long range interactions appear with the fluctuations  of the linking number: it 
is known
experimentally that the energy cost  associated with
the residual linking number for supercoiled DNA
is given by
$$G_r= \frac{K\alpha_r^2}{2}=\frac{K}{2}(\alpha+\frac{n}{A}-{\cal T})^2.$$
Experimentally, the coefficient $K$ is inverse proportional to the number
of bases of the DNA; we thus set
$$K=\frac{K_0}{N}.$$
As $N$ is large, basic statistical reasoning suggests to renormalize
the variable $n$ as $n/N$, to catch the thermodynamical limit.
We thus introduce the superhelical density $\kappa$ by setting
$$\alpha=\kappa N.$$
The overall free energy takes then the form
$$G=\frac{C}{2}\sum_{i=1}^N 
n_i\tau_i^2+N\frac{K_0}{2}\big(\kappa+\frac{n}{2NA}-\frac{1}{N}\sum_{i=1}^N 
\frac{n_i\tau_i}{2\pi}\big)^2
+\sum_{i=1}^N ((a+2b_i)n_i-an_{i}n_{i+1}).$$
 We use the local fields $2b_i$ insteed of $b_i$
 for notational purpose.

\section{Results on denaturation}\label{s.denaturation}

\subsection{The homopolymer approximation}
\label{s.homo}

In this paragraph, we suppose that $b_i\equiv b>0$ and
that $\tau_i\equiv \tau \in \R$.  Set for convenience
$$M_N=\frac{1}{N}\sum_{i=1}^N n_i,\ m_N=\frac{1}{N}\sum_{i=1}^N\sigma_i=2M_N-1.$$
Then the Hamiltonian of the system becomes
$$G_\tau=
N\big(2bM_N+\frac{C\tau^2}{2}M_N+\frac{K_0}{2}\big(\kappa+(\frac{1}{2A}-\frac{\tau}{2\pi})M_N\big)^2\big)
+a H_{\rm sing}/4,$$
where the index $\tau$ of $G_\tau$ indicates the dependence on the torsion 
coefficient $\tau$ and
$H_{\rm sing}$ denotes the Hamiltonian associated with the nearest neighbor 
ferromagnetic Ising model in
dimension 1
$$H_{\rm sing}=-\sum_{i=1}^{N-1}\sigma_i \sigma_{i+1}=4r-N.$$
Before introducing the Gibbs measure of the system at inverse temperature
$\beta >0$, let us proceed as in Benham by averaging the system
with respect to the torsion coefficient $\tau$. The Boltzman weight
should be
$\exp(-\beta G_\tau) $. Averaging over $\tau \in\R$ gives the
integral $\int_\R \exp(-\beta G_\tau)d\tau$, that is the effective Hamiltonian
\begin{equation}\label{EffectHamil}
H_1=
 N 2bM_N+\frac{a}{4} H_{\rm sing} +N \frac{2\pi^2C K_0}{4\pi^2C+K_0 M_N}\big(\kappa
+\frac{M_N}{A}\big)^2,
 \end{equation}
 (see Fye and Benham(1999)).
 


 \subsubsection{Arbitrary large perimeter}\label{Laplace}

 In this approximation, we shall consider the behavior of $M_N$ in the
thermodynamical limit $N\to\infty$
 under Gibbs measure
 \begin{equation}\label{GibbsHomo}
 \piB(\sigma)=
 \exp(-\beta H_1(\sigma))/Z_N(\beta,B),
 \end{equation}
 where $Z_N(\beta,B)$ denotes the related partition function,
 and where $B=(b_i)_{1\le i\le N}$, $b_i\equiv b$, denotes the exterior field.
  We use mainly Laplace method
 bu using the large deviation rate function associated with the ferromagnetic
 nearest neighbor Ising model in dimension one.     Notice the appearance
 of the magnetization $M$ in the denominator   of (\ref{EffectHamil}),
  which is quite
 unconventional.   Let $\pi_{\beta_a}$ be the Gibbs measure at inverse 
temperature
 $\beta_a=a\beta/4$ associated with the Hamiltonian $H_{\rm sing}$. Then, for
 any function $h:I:=[0,1]\longrightarrow\R$,
 \begin{equation}\label{Mean}
 <h(M_N)>_{\piB}=\frac{<h(M_N)\exp(-N\beta F_B(M_N))>_{\pi_{\beta_a}}}{<\exp(-N\beta
F_B(M_N))>_{\pi_{\beta_a}}},
 \end{equation}
 where
 \begin{equation}\label{F}
 F_B(y)=b(2y-1)+ G(y),\ y\in {\rm I},
 \end{equation}
 and
 $$G(y)=\frac{2\pi^2C K_0}{4\pi^2C+K_0 y}\big(\kappa +\frac{y}{A}\big)^2.$$
 Let $\mu_N$ be the law of $m_N(\sigma)=2M_N(\sigma)-1$ under Gibbs measure
 $\pi_{\beta_a}$. Then
 \begin{equation}\label{Ratio}
   <h(M_N)>_{\piB}=
 \frac{\int_{-1}^{+1}\mu_N({\rm dz})h(\frac{1+z}{2})\exp(-\beta N 
F_B(\frac{1+z}{2}))}{ \int_{-1}^{+1}\mu_N({\rm dz})\exp(-\beta N 
F_B(\frac{1+z}{2}))}.
 \end{equation}
 Benham(1979) investigates the thermodynamics
 of supercoiled DNA, by minimizing free energies,
 and introduces critical thresholds of supercoiling.
 In this macroscopic approach, it is shown that
 sufficient negative
 supercoiling implies local denaturation.
 In Benham(1996), this work is extended to positively supercoiled DNA 
($\kappa>0$);
  looking at the various plots contained in this work, we see the appearance
  of critical superhelical densities
 above  which
 a positively supercoiled DNA remains intact
 at temperatures higher than the melting point. Similarly, we
 introduce the
 \begin{definition}\label{Phase}
The order parameter of the system is
$$<M_N>_{\piB},\ \hbox{ or }<m_N>_{\piB},$$
the magnetization of the spin system.
We say that the system exhibits phase
transitions when there
exist critical superhelical densities
$\bar\kappa_c(B)>\kappa_c(B)$ such that,
in the large $N$ limit,
$<M_N>_{\piB}\to 0$ as $\kappa> \bar\kappa_c(B)$,
$<M_N>_{\piB}\to 1$ as $\kappa<\kappa_c(B)$,
and $\lim <M_N>_{\piB}\in (0,1)$ otherwise.
\end{definition}

 Let $\Lambda_N(\lambda)$ be the logarithmic  moment generating function
 $$\Lambda_N(\lambda):=\ln(\pi_{\beta_a}(\exp(Nm_N\lambda))),$$
 with (see e.g. Baxter(1982), p.34)
 $$\Lambda_{\infty}(\lambda):=\lim_{N\to\infty} 
\frac{1}{N}\Lambda_N(\lambda)=\ln\big(\frac{e^{\beta_a}\cosh(\lambda)+\sqrt{e^{2\beta_a}\sinh(\lambda)^2+e^{-2\beta_a}}}{e^{\beta_a}+e^{-\beta_a}}).$$
Then,
according to Gaertner-Ellis Theorem, (see e.g. Dembo and Zeitouni(1992))
the law of $m_N$ under $\pi_{\beta_a}$ satisfies a large deviation principle with
good rate function
$$I_{\rm sing}(z)=\sup_{\lambda\in\R}(\lambda z-\Lambda_{\infty}(\lambda)),$$
the Legendre transform of $\Lambda_\infty$. $\Lambda_\infty'=e^{\beta_a}
\sinh/\sqrt{e^{2\beta_a}\sinh^2+e^{-2\beta_a}}$, and
$\Lambda_\infty''=e^{\beta_a} \cosh/(e^{2\beta_a}\sinh^2+e^{-2\beta_a})$. 
$\Lambda_\infty$
is thus strictly convex on ${\bf R}$, and its Legendre transform is
essentially smooth (see Theorem 26.3 in Rockafellar(1972)). The derivative $I_{\rm sing}'(z)$ tends to $+\infty$
when $z$ converges to a boundary point of the domain of $I_{\rm sing}$. 
From computation,
\begin{eqnarray*}
I_{\rm sing}(z)&=&z\ln(\frac{z 
e^{-2\beta}+\sqrt{1+z^2(e^{-4\beta_a}-1)}}{\sqrt{1-z^2}})\\
&\ &\ \ 
-\ln(\frac{e^{\beta_a}\sqrt{1+z^2(e^{-4\beta_a}-1)}+e^{-\beta_a}}{\sqrt{1-z^2}(e^{\beta_a}+e^{-\beta_a})}),
\end{eqnarray*}
 $\vert z\vert \le 1$,
and $I_{\rm sing}(z)=+\infty$ when $\vert z\vert > 1$.

In the special case where $a=0$, the rate function becomes the entropy
$$I_{\rm sing}(z)=\frac{1+z}{2}\ln(1+z)+\frac{1-z}{2}\ln(1-z),\ \vert z\vert\le 
1.$$
The integrals appearing in (\ref{Ratio}) can be estimated through Laplace's
method by rewritting the numerator
heuristically  as
$$\int_{-1}^{+1}{\rm dz}h(\frac{1+z}{2})\exp(-N J(z)),$$
where we set
$$J(z):= I_{\rm sing}(z)+\beta F_B(\frac{1+z}{2}),\ \vert z\vert\le 1.$$
Then (\ref{Ratio}) is asymptotically equivalent to
$$\frac{\int_{-1}^{+1}{\rm dz} h(\frac{1+z}{2})\exp(-N (J(z)-\inf_{\vert 
z\vert\le 1}J(z)))}{\int_{-1}^{+1}{\rm dz} \exp(-N (J(z)-\inf_{\vert z\vert\le 
1}J(z)))}.$$

\begin{theorem}\label{Order}
 $\exists$ a unique $z_*\in\ (-1,+1)$ minimizing
 $J(z)$, with
 $$<M_N>_{\pi_{\beta,B}}\ \longrightarrow \frac{1+z_*}{2}.$$
\end{theorem}

    The model does not exhibit phase transitions
    in the sense of Definition \ref{Phase}. This is
    a consequence of the stiffness of the rate function
     $I_{\rm sing}(z)$ as $\vert z\vert\to 1$:
    $\lim_{\vert z\vert\to 1}({\rm d}/{\rm dz})I_{\rm sing}(z)= +\infty$,
    and the rate function $J(z)$ can not be decreasing
    in the neighborhood of $z=1$. The minima of $J(z)$ are thus
    located in the interior of the unit interval. Suppose
    that the Ising measure is replaced by some probability
    measure $V_N$ on $\Omega_N$, such that the law of $M_N$
    under $V_N$ satisfies a large deviation principle
    with strictly convex and smooth free energy function
    $\Lambda$; then its Legendre transform is essentially
    smooth, and again, using Varadhan's Theorem, one gets the rate function
    $J(z)-\inf_z J(z)$; similarly, the minima of
    $J$ are located in the interior of the domain
    of the Legendre transform, and no phase transition
    occurs.
 \bigskip

\noindent {\it Proof}: Consider the probability measure
$$\nu_N(C)=\frac{\int_C \mu_N({\rm d}z)\exp(-N\beta F_B((1+z)/2)}{\int_{-1}^1 
\mu_N({\rm d}z)\exp(-N\beta F_B((1+z)/2)},$$
for any Borel subset $C$ of $[-1,1]$. Varadhan's Theorem
(see Deuschel and Stroock, Theorem 2.1.10 and
    exercice 2.1.24) gives that
    $\nu_N$ satisfies a large deviation principle with
    good rate function $J(z)$. If $J$ attains its infimum
    at a unique point $z_*$ of the interval, the sequence
    $\nu_N$ converges weakly to the point mass $\delta_{z_*}$.
    Consider first $F_B(y)$, $y\in [0,1]$, or equivalently
$F_B +2b\kappa A+b$, which is equal to
$$\frac{2(bA^2+\pi^2 C)}{A^2}\frac{(y-y_0)(y-y_1)}{(y-y_2)},$$
where
$$y_0=-\kappa A >0,\ y_1=-\frac{\pi^2 C A}{K_0(b A^2+\pi^2C)}(4bA+K_0\kappa)\hbox{ and }y_2=-\frac{4\pi^2 C}{K_0}.$$
Notice that $b>0$ implies that $y_1>y_2$. The function has
a pole at $y=y_2<0$, and two roots $y_1$ and $y_0>0$.
Then $F_B$ is strictly convex on the half line
$(y_2,+\infty)$. Thus $J$ is strictly convex on 
$[-1,1]$, with $\lim_{\vert z\vert\to 1}J'(z)=+\infty$.
The unique infimum is thus located in the interior of the
interval.



    \subsubsection{Limited perimeter}
    \label{s.s.peri}

Computations done in some
    theoretical studies (see e.g. Benham(1989), p. 268 and Benham(1990), p. 6302)
    or empirical studies( see e.g. Sun {\it et altri}(1995, p. 8658))
    deal with the behavior of $M_N$ when
 $2r$ is fixed, or is small.  In what follows, we give
conditions on
    the growth of $r=r_N$ ensuring the possibility
    of phase transitions, that is the possibility for the
    existence of a stable and robust denatured state
    when the superhelical density is small enough.

    In what follows, we condition on the event $\{\sigma;\ \vert\sigma\vert = 
2r_N\}$, where
    for any configuration $\sigma$, $\vert\sigma\vert$ denotes the perimeter of 
$\sigma$,
    with $\vert\sigma\vert=\vert i;\ \sigma_i=-\sigma_{i+1}\vert=2r$, and
    $H_{\rm sing}(\sigma)=2\vert\sigma\vert -N$.
     Classical combinatorics (see e.g. Feller(1971)) shows that
    the number of configurations of length $N$ with $\sum_{i=1}^N n_i=n$ and
    perimeter $1\le r\le N/2$ is given by
    $$M(n,r)=\frac{N}{r}{n-1\choose r-1}{N-n-1\choose r-1}.$$
    Let $U_{r,N}$ be the uniform probability measure
    on the subset ${\cal C}_{r,N}$ of the cube
    $\Omega_N$ consisting of spins of perimeter $2r$.
    Let $P_{r,N}$ be the law of $M_N$ under $U_{r,N}$,
    with
    ${\rm supp}(P_{r,N})=\{r/N,\cdots,1-r/N\}$, given by
    \begin{equation}\label{Proba}
    P_{r,N}(n/N)=\frac{{n-1\choose r-1}{N-n-1\choose r-1}}{\sum_{r\le n\le N-r} 
{n-1\choose r-1}{N-n-1\choose r-1}}.
    \end{equation}
   Then the average $<h(M)>_{\piB}$ becomes
   $$
   \frac{\sum_r \exp(-\beta_a r)\vert\{\vert\sigma\vert=2r\}\vert\sum_{r\le n\le
N-r}P_{r,N}(n/N)h(n/N)\exp(-N\beta F_B(n/N))}{ \sum_r \exp(-\beta_a
r)\vert\{\vert\sigma\vert=2r\}\vert\sum_{r\le n\le N-r}P_{r,N}(n/N)\exp(-N\beta 
F_B(n/N))}$$
   We will be concerned with integrals of the form
   $$\int_0^1 \P({\rm d}y) h(y)\exp(-\beta N F_B(y)),$$
   when $r=r_N$ is such  that $r_N/N\longrightarrow 0$ as $N\to\infty$.

    \begin{lemma}\label{Zero}
    Assume that $2r_N<N$ and that $r_N/N\longrightarrow 0$
    as $N\to\infty$. Then
    \begin{equation}\label{Condition}
    \frac{1}{N}\log{n\choose r_N}\longrightarrow 0,\ N\to\infty,
    \end{equation}
    when $n=[\rho N]$, for $0<\rho\le 1$.
    The sequence of probability measures $(P_{r_N,N})_{n\in {\bf N}}$
    satisfies a large deviation principle with good rate function
    $I^r:{\bf R}\longrightarrow [0,+\infty)$ given by
    $I^r(y)=0$, $y\in I$, and $I^r(y)=+\infty$, $y\in I^c$.
    \end{lemma}
     {\it Proof}: The first assertion is a consequence of Stirling's formula.
     Assume that $A=(a,b)\subset I$.
       For $N$ large enough,
       $A\cap\{r_N/N,\cdots,1-r_N/N\}\ne\emptyset$, and $A$ contains
       an element $\rho_N$ of the form
       $\rho_N=[\rho N]/N$ for some $0<\rho<1$, with
 $$\log\P(\{\rho_N\})\le \log\P(A)\le 0.$$
 Using (\ref{Proba}) and (\ref{Condition}),
     it remains to check that
     $$\frac{1}{N}\log( \sum_{r\le n\le N-r}{n\choose r}{N-n\choose r} 
)\longrightarrow 0.$$
     But, as the sum larger than one,
     $$
     0\le \frac{1}{N}\log( \sum_{r\le n\le N-r}{n\choose r}{N-n\choose r} )$$
     $$\le\frac{1}{N}\log((N-2r)\sup_{r\le n\le N-r}{n\choose r}{N-n\choose 
r})$$
     $$\le\frac{\log(N)}{N}+\frac{1}{N}\log(\sup_{r\le n\le N-r}{n\choose 
r}{N-n\choose r}).  $$
     The sequence $({n\choose r}{N-n\choose r})_n$ attains its supremum when
     $n=[N/2]$, and the statement is a consequence of (\ref{Condition}).
     $A$ is a $I^r$-continuity set. When $A\cap I=\emptyset$,
     $P_{r_N,N}(A)\equiv 0$, which is consistent with
     $I^r(y)=+\infty$ when $y\not\in I$. When $A$ takes the form
     $A=[-c,\varepsilon]$ with $c>0$ and $\varepsilon >0$, $\exists$
     $N_0\in {\bf N}$ such that $0<r_N/N<\varepsilon$, $\forall N\ge N_0$,
     and the same argument applies. When $A=[-c,0]$, $P_{r_N,N}(A)\equiv 0$,
     and we have the inequalities defining the large deviation principle
     $$-\inf_{y\in A^0}I^r(y)\le\liminf \frac{1}{N}\log(P_{r_N,N}(A))$$
     and
     \begin{equation}\label{LargeDeviation}
     \limsup \frac{1}{N}\log(P_{r_N,N}(A))\le -\inf_{y\in\bar A}I^r(y).
     \end{equation}
      The above arguments show that the upper and lower bounds
      (\ref{LargeDeviation}) hold for open and compact sets.
      The sequence of measures $(P_{r_N,N})$ is supported
      by the unit interval, and the sequence is exponentially tight.
      The large deviation principle follows.

     \medskip

     When   the large deviation principle is satisfied
     with the flat rate function $I^r$,  Laplace's method gives
     that the system exhibits a phase transition with respect
     to the order parameter $M_N$:  the mass of the integral
     is located near the infimum of the function $F_B(y)$, and therefore,
     the DNA is completely denatured when the parameters of
     the problem are such that
     $$\inf_{y\in (0,1)}F_B(y)=F_B(1).$$
     This occurs for example when the helicity density $\kappa$ is smaller
     than a critical density $\kappa_c$.

\begin{theorem}\label{CriticalTransitions}
The function $F_B:I\longrightarrow {\bf R}$ 
attains its infimum at a unique point
$y_*(B,\kappa)$.
Let
$$y_0=-\kappa A >0,\ y_1=-\frac{\pi^2 C A}{K_0(b A^2+\pi^2
C)}(4bA+K_0\kappa)\hbox{ and }
y_2=-\frac{4\pi^2 C}{K_0}.$$
Let $\bar\kappa_c(B)$ and $\kappa_c(B)$
be the smallest roots of the polynomials
$P(\kappa)=(y_0-y_2)(y_1-y_2)-y_2^2$
and
$Q(\kappa)=P(\kappa)-(1-2y_2)$,
with
$\bar\kappa_c(B)>\kappa_c(B)$.
Set
$M_*=y_2+\sqrt{(y_2-y_0)(y_2-y_1)}$.
Then
i) $\kappa>\bar\kappa_c(B)$ implies that
$F_B$ is increasing on ${\rm I}$ with
$y_*(B,\kappa)=0$, ii) $\kappa_c(B)<\kappa<\bar\kappa_c(B)$
implies that
$y_*(B,\kappa)=M_*\in (0,1)$,
with $F_B$ decreasing below $y_*(B,\kappa)$ and increasing
above $y_*(B,\kappa)$,
 and
iii) $\kappa<\kappa_c(B)$ implies that
$F_B$ is decreasing on ${\rm I}$ with
$y_*(B,\kappa)=1$.
Let $h:{\rm I}\longrightarrow {\bf R}$
be bounded and continuous.
 Assume that
$r_N/N\longrightarrow 0$ as $N\to\infty$. Then
$$\frac{P_{r_N,N}(h\exp(-N \beta F_B))}{P_{r_N,N}(\exp(-N \beta 
F_B))}\longrightarrow h(y_*(B,\kappa)).$$
\end{theorem}
{\it Proof:} We look for the infimum of $F_B$
on I; equivalently, we can consider the infimum
of $F_B +2b\kappa A+b$, which is equal to
$$\frac{(2bA^2+2\pi^2 C)}{A^2}\frac{(y-y_0)(y-y_1)}{(y-y_2)}.$$
Notice that $b>0$ implies that $y_1>y_2$. The function has
a pole at $y=y_2<0$, and two roots $y_1$ and $y_0>0$. We see
that the first part of the theorem (cases i), ii) and iii))
is related to the location of the infimum of the function
with respect to $(y_2,0)$, $(0,1)$
and $(1,\infty)$. Taking the derivative,
we must look for the roots of
$y^2-2y_2y+y_2(y_0+y_1)-y_0y_1$,
of discriminant
$(y_0-y_2)(y_1-y_2)\ge 0$, since
$y_1>y_2$, $y_0>0$ and $y_2<0$, and we get the condition
for the largest root $M_*$. The polynomial $P(\kappa)$
is obtained
by imposing $M_*<0$, and $Q(\kappa)$
by imposing $M_*<1$.
Concerning the last assertion, consider the 
probability measure
$$\mu_N(A):=\frac{\int_A \exp(-N\beta F_B(y))P_{r_N,N}({\rm d}y)}{\int_I 
\exp(-N\beta F_B(y)) P_{r_N,N}({\rm d}y)},$$
for any Borel subset $A$.
From Varadhan's Theorem (see e.g. 
Theorem 2.1.10 and exercice 2.1.24 in Deuschel and Stroock(1989) or
Theorem II.7.2 in Ellis(1985)) and Lemma
\ref{Zero},
the sequence $(\mu_N)$ satisfies a large deviation
principle with rate function $I_F(y)-\inf_y I_F(y)$, where $I_F(y)=I^r(y)+\beta 
F_B(y)$.
$\inf_{y\in {\bf R}}I_F(y)=\beta \inf_{y\in {\rm I}}F_B(y)$,
which is realized at a unique element $y_*(B,\kappa)$;
$\mu_N$ converges then weakly to the point mass $\delta_{y_*(B,\kappa)}$.




 \subsection{The heteropolymer Case}\label{Hetero}

In the heteropolymer case, the field $B^N=(b_i)_{1\le i\le N}$
is indexed by a word of length $N$ on the two letters alphabet
$\{A+T, G+C\}$. 
Given $B^N$, let
$$\rho^+_N(B^N):=\frac{1}{N}\sum_{i=1}^N {\rm I}(b_i=b_{AT}),$$
where ${\rm I}(\cdot)$ is the indicator function. Given some proportion
$\rho^+\in {\rm I}$, we shall consider families $(B^N)_{N\in {\bf N}}$
of words with $\rho^+_N(B^N)\longrightarrow \rho^+$. 
Let
$$m_N^+(\sigma)=\frac{1}{N}\sum_{i=1}^N {\rm I}(b_i=b_{AT})\sigma_i
\hbox{ and }m_N^-(\sigma)=\frac{1}{N}\sum_{i=1}^N {\rm I}(b_i=b_{GC})\sigma_i,$$
with $m_N=m_N^+ +m_N^-$, and
consider
the mapping $\Psi_{B^N}:\Omega_N\longrightarrow {\bf R}^2$
given by $\Psi_{B^N}(\sigma)=(m_N^+(\sigma),m_N^-(\sigma))$,
which permits to control the eventual localization
of the magnetization (see Mathieu and Picco(1998)
for other use of this mapping in random fields
Curie Weiss models).
Let
$V_N$ be a probability measure on $\Omega_N$, and let
$Q_N$ be the image measure of $V_N$ under $\Psi_{B^N}$.
We consider 
the behavior of $m_N^+$ and $m_N^-$ under
the Gibbs measure
$$\pi_{\beta,B^N,V_N}(\sigma)=\frac{V_N(\sigma)\exp(-N\beta 
F_{B^N}(\sigma))}{Z_N(\beta,B^N,V_N)},$$
where we set
$$F_{B^N}(\sigma)=G(M_N(\sigma))+\frac{1}{N}\sum_{i=1}^N b_i \sigma_i.$$
Let us denote by $B_{GC}$ and $B_{AT}$ the fields associated
with the GC and AT homopolymers, with critical superhelical
densities $\bar\kappa_c(B_{GC})$, $\kappa_c(B_{GC})$,
$\kappa_c(B_{AT})$ and limiting proportion of broken bonds
$y_*(B_{GC},\kappa)$
(see Theorem \ref{CriticalTransitions}).
\begin{theorem}\label{Proportion}
Let $(B^N)_{N\in {\bf N}}$ be a sequence of words
of $\{b_{AT},b_{GC}\}^N$ with
$\rho^+_N(B^N)\longrightarrow \rho^+$, for
some $\rho^+\in {\rm I}$.
Suppose that the family of probability measures
$Q_N$ satisfies a large deviation principle
with  rate function $I_D:{\bf R}^2\longrightarrow [0,+\infty)$
given by
$I_D(y)=0$, $y\in D$ and $I_D(y)=+\infty$, $y\in D^c$,
where
$D=[-\rho^+,\rho^+]\ {\rm x}\ [-\rho^-,\rho^-]$.
Let
$\kappa$ be such that $\kappa_c(B_{GC})<\kappa<\kappa_c(B_{AT})$.
Assume that $\rho^+>y_*(B_{GC},\kappa)$: Then
denaturation is localized on the AT domain, that is
$<m_N^+>_{\pi_{\beta,B^N,V_N}}\longrightarrow \rho^+$,
and
$<m_N^->_{\pi_{\beta,B^N,V_N}}\longrightarrow -\rho^-$,
where we set $\rho^-=1-\rho^+$, the proportion of GC bonds.
Conversely, assume that $\rho^+<y_*(B_{GC},\kappa)$: Then
$<m_N^+>_{\pi_{\beta,B^N,V_N}}\longrightarrow \rho^+$,
and
$<m_N^->_{\pi_{\beta,B^N,V_N}}\longrightarrow
2y_*(B_{GC},\kappa)-1-\rho^+>-\rho^-$.
\end{theorem}
\begin{remark}
If the parameters of the problem are such that
$\bar\kappa_c(B_{GC})<\kappa_c(B_{AT})$
(recall that the fields are temperature dependent, see (\ref{FreeEnergy})), and
$\kappa$ is so that $\bar\kappa_c(B_{GC})<\kappa<\kappa_c(B_{AT})$,
denaturation is localized on the AT domain,
for arbitrary proportion $\rho^+$ of AT bonds.
However, when $\kappa<\bar\kappa_c(B_{GC})$, denaturation is
localized on the AT domain when $\rho^+>y_*(B_{GC},\kappa)$
and denaturation expands beyond the AT domain when
$\rho^+<y_*(B_{GC},\kappa)$.
\end{remark}
{\it Proof:} We must evaluate the asymptotic behavior of
$$\frac{\sum_{\sigma}V_N(\sigma)\exp(-N\beta
F_{B^N}(\sigma))h(m_N^+(\sigma),m_N^-(\sigma))}{\sum_{\sigma}V_N(\sigma)\exp(-N\beta
F_{B^N}(\sigma))},$$
 for functions $h$ of the two variables $(m_N^+,m_N^-)$. Notice that
 \begin{eqnarray*}
 F_{B^N}(\sigma)&=&G(M_N(\sigma))+b_{AT}m_N^+(\sigma)+b_{GC}m_N^-(\sigma)\\
     &=&G(M_N(\sigma))+b_{GC}m_N(\sigma)+(b_{AT}-b_{GC})m_N^+(\sigma),
     \end{eqnarray*}
 Let
 $F(m^+,m^-)=G(M)+b_{AT}m^++b_{GC}m^-$, where $m=m^+ + m^-$ and
 $M=(m+1)/2$. We must thus check the behavior of
 the expectation
 $$\mu_N(h)=\frac{\int h(y) \exp(-N\beta F(y))Q_N({\rm dy})}{\int \exp(-N\beta 
F(y))Q_N({\rm dy})},$$
 for bounded and continuous functions $h$. From Varadhan's Theorem,
 the family of probability measures $\mu_N$ satisfies a large deviation
 principle with rate function $I_F(y)-\inf_y I_F(y)$, where $I_F(y)=I_D(y)+\beta 
F(y)$.
 $\inf_y I_F(y)=\beta\inf_{y\in D}F(y)$. If this infimum
 is realized at a unique point $y_*$ of $D$, the sequence of measures
 $\mu_N$ converges weakly to the Dirac mass $\delta_{y_*}$.
 We thus look for the minima of $F$ on $D$.
$$\inf_{(m^+,m^-)\in D} F(m^+,m^-)$$
$$=\inf_{\vert m\vert\le 1} (G(M)+b_{GC}m +(b_{AT}-b_{GC})\sup_{(m^+,m^-):\ 
m^-+m^+=m}m^+)$$
Given, $\rho^+$, consider $m$ such that
$M=(m+1)/2\ge \rho^+$, that is 
$m\ge 2\rho^+-1$. Then
$\sup_{m^+ + m^-=m}m^+=\rho^+$(corresponding to (\ref{A}) below)): when $m^+=\rho^+$,
one obtains $m^-=m-\rho^+$ and the pair
$(m^+,m^-)$ is element of $D$ since
$m^-=m-\rho^+\le\rho^-$ if and only if
$m\le \rho^+ +\rho^-=1$ and
$m^-\ge -\rho^-$ if and only if
$m\ge \rho^+ -\rho^-=2\rho^+-1$
 Similarly,
when $m$ is such that $M\le \rho^+$, the maximal
possible value of $m^+$ is $m^+=2M-\rho^+$ (corresponding to (\ref{B}) below).
In summary
the infimum is obtained by taking the minimum between
\begin{equation}\label{A}
\inf_{M\ge\rho^+}G(M)+b_{GC}m+(b_{AT}-b_{GC})\rho^+,
\end{equation}
and
\begin{equation}\label{B}
\inf_{M<\rho^+}G(M)+b_{AT}m +\rho^-(b_{AT}-b_{GC}).
\end{equation}
We next use the hypotheses.  From Theorem \ref{CriticalTransitions},
$\kappa<\kappa_c(B_{AT})$ implies that 
$G(M)+b_{AT}m$ attains its infimum when
$m_*=1$, or $M_*=1$, and is
decreasing on the unit interval;
 (\ref{B}) becomes
$$G(\rho^+)+b_{AT}(2\rho^+-1)+(1-\rho^+)(b_{AT}-b_{GC})=
G(\rho^+)+\rho^+ b_{AT}-\rho^- b_{GC}.$$
Concerning (\ref{A}), $\kappa>\kappa_c(B_{GC})$ and,
from Theorem \ref{CriticalTransitions}, the function
$G(M)+b_{GC}m$ attains its minimum at
$y_*(B_{GC},\kappa)$, and is increasing above
this point. Thus
$$\inf_{M\ge\rho^+}G(M)+b_{GC}m=G(\rho^+)+(2\rho^+-1)b_{GC},$$
when $\rho^+\ge y_*(B_{GC},\kappa)$. Then, both (\ref{A})
and (\ref{B}) are minimized for $M=\rho^+$, which is
the maximal value of $m^+$. Thus
$m^+=\rho^+$ and $m^-=m-m^+=2\rho^+-1-\rho^+=-\rho^-$,
as required.

\noindent Conversely,
assume that $\rho^+<y_*(B_{GC},\kappa)$. Then (\ref{A}) becomes
$$\inf_{M\ge\rho^+}G(M)+b_{GC}m+(b_{AT}-b_{GC})\rho^+$$
$$=G(y_*(B_{GC},\kappa))+b_{GC}(2y_*(B_{GC},\kappa)-1)+(b_{AT}-b_{GC})\rho^+.$$
When $M<\rho^+$, the infimum is still given by
$$G(\rho^+)+b_{AT}\rho^+-\rho^- b_{GC}=G(\rho^+)+b_{GC}(2\rho^+-1)+\rho^+ 
b_{AT}-\rho^+ b_{GC},$$
and one obtains that the minimum is realized when $M=y_*(B_{GC},\kappa)$, and 
therefore
$m^+=\rho^+$ and $m^-=m-m^+=2M-1-m^+=2y_*(B_{GC},\kappa)-1-\rho^+$,
as required.
\bigskip

In the remaining, we give an example of probability measure
$V_N$ on $\Omega_N$ such that 
$Q_N=V_N\circ \Psi_{B^N}^{-1}$ satisfies a large
deviation principle with rate function $I_D$.
Given a sequence $(B^N)$, consider the
family of spins $\sigma^N$ given by
$\sigma^N_i={\rm I}(b^N_i=b_{AT})-{\rm I}(b^N_i=b_{GC})$.
Given a sequence $r_N$, consider the restricted
Ising measure
$$V_N(\sigma)=\pi_{\beta_a,r_N}(\sigma)=\frac{{\rm I}(\vert\sigma\vert\le 
2r_N)\pi_{\beta_a}(\sigma)}{
\sum_{\sigma\in\Omega_N}{\rm I}(\vert\sigma\vert\le 
2r_N)\pi_{\beta_a}(\sigma)}.$$
\begin{lemma}\label{FlatRate}
Let $(B^N)_{N\in {\bf N}}$ be a sequence of
words such that $\rho_N^+\longrightarrow \rho^+\in {\rm I}$,
and $\sigma^N\in {\cal C}_{\bar r_N,N}$, for some
sequence $(\bar r_N)_{N\in {\bf N}}$. Assume that
$\bar r_N+2\le r_N$, and that $r_N/N\longrightarrow 0$
as $N\to\infty$. Then the probability measure
$Q_N=\pi_{\beta_a,r_N}\circ \Psi_{B^N}^{-1}$
satisfies a large deviation principle with rate
function $I_D$.
\end{lemma}
This is an extension of the homopolymer case:
$\bar r_N/N\to 0$ means that the word
associated with the DNA is formed of relatively
large droplets of A+T bonds alternating with
similar droplets of G+C bonds.

\noindent {\it Proof:}
When $r_N/N\to 0$, (\ref{Condition})
implies that
$\log(V_N(\tilde\sigma^N))/N\to 0$ for any
sequence of spins $(\tilde\sigma^N)_N$ 
with $\vert\tilde\sigma^N\vert\le 2 r_N$.
Now, given an open subset $A$ of $D^0$,
containing some point
$\lambda=(\lambda^+,\lambda^-)$,
with $\vert\lambda^+\vert<\rho^+$ and
$\vert\lambda^-\vert<\rho^-$, consider
the sequence $\lambda_N=([N\lambda^+]/N,[N\lambda^-]/N)$,
which is in $A$ for $N\ge N_0(A,\lambda)$.
Then 
\begin{equation}\label{Inverse}
\exists \tilde\sigma^N\in {\cal C}_{\tilde r_N,N}
\hbox{ with }\tilde r_N\le \bar r_N+1\hbox{ and 
}\Psi_{B^N}(\tilde\sigma^N)=\lambda_N,\
\forall N\ge N_0(A,\lambda).
\end{equation}
Suppose that (\ref{Inverse}) is true: Then one obtains
$$V_N(\tilde\sigma^N)\le Q_N(\{\lambda_N\})\le Q_N(A)\le 1,$$
and it follows that $\log(Q_N(A))/N\to 0$ as $N\to\infty$.
In this case, $A$ is a $I_D$-continuity set. The main
property to check is thus (\ref{Inverse}).
Set $\lambda_N^{\pm}=[\lambda^{\pm}N]/N$, and define
$M_N^\pm = (\lambda_N^\pm+\rho_N^\pm)/2$, where
$\rho_N^-=1-\rho_N^+$. Choose an origin in the 
circular DNA at some site $i_0$ with
$\sigma_{i_0-1}^N=-1$ and $\sigma_{i_0}^N=1$.
We can consider the linear string, starting at 
$i_0$, with an A+T droplet, and ending with a
G+C droplet. Let $A_1,\cdots,A_{\bar r_N}$ be
the A+T droplets, ordered according to their appearance
along the string, and define similarly 
$G_1,\cdots,G_{\bar r_N}$. Set
$a_j=\vert A_j\vert$ and $g_j=\vert G_j\vert$,
$1\le j\le \bar r_N$. The string is viewed as
the juxtaposition of symbols $A_1G_1\cdots A_{\bar r_N}G_{\bar r_N}$.
Let $T^+$ and $T^-$ be defined by
$$T^+=\min\{1\le j\le \bar r_N;\ \sum_{k=1}^j a_k\ge N M_N^+\},$$
and
$$T^-=\min\{1\le j\le \bar r_N;\ \sum_{k=1}^j g_k\ge N M_N^-\}.$$
Notice that both $T^+$ and $T^-$ are well defined for $N$
large enough since $\sum_{k=1}^{\bar r_N}a_k=N\rho_N^+$,
$\sum_{k=1}^{\bar r_N}g_k=N\rho_N^-$, $\rho_N^+\to \rho^+$,
$\rho_N^-\to \rho^-$, $M_N^+\to (\lambda^+ +\rho^+)/2<\rho^+$
and $M_N^-\to (\lambda^- +\rho^-)/2<\rho^-$. Let $i_1$ be the
site situated in $A_{T^+}$ such that
$$\vert\{i\in\ A_1\cup \cdots\cup A_{T^+};\ i\le i_1\}\vert = N M_N^+,$$
and define similarly $i_2$ for the G+C domain.
Set
$\tilde \sigma^N_i=+1$ when $i\in A_1\cup\cdots\cup A_{T^+}$ and
$i\le i_1$, $\tilde \sigma^N_i=-1$ when $i\in A_{T^+}\cup\cdots\cup A_{\bar 
r_N}$ and
$i> i_1$,
$\tilde \sigma^N_i=-1$ when $i\in G_1\cup\cdots\cup G_{T^-}$, $i\le i_2$,
and $\tilde\sigma^N_i=+1$ when
$i\in G_{T^-}\cup\cdots\cup G_{\bar r_N}$, $i>i_2$. Clearly
\begin{eqnarray*}
\sum_{i\in \cup_{1\le k\le\bar r_N}A_k}\tilde\sigma^N_i&=& N M_N^+-(N\rho_N^+ - 
N M_N^+)\\
               &=&N(2M_N^+-\rho_N^+)=N\lambda_N^+,
\end{eqnarray*}
and similarly 
$$\sum_{i\in \cup_{1\le k\le \bar r_N}G_k}\tilde\sigma^N_i=N \lambda_N^-,$$
giving $\Psi_{B^N}(\tilde\sigma^N)=\lambda_N$, as required. Next suppose
without loss of generality that $T^+\le T^-$, then $i_1<i_2$, and,
from construction, $\tilde\sigma^N_i=\sigma^N_i$ when
$i\le i_1$, $\tilde\sigma^N_i=-\sigma^N_i$ when
$i>i_2$ and $\tilde\sigma^N_i=-1$ when
$i_1<i\le i_2$. It follows that $\vert\tilde\sigma^N\vert\le 2(\bar r_N+1)$,
as required.

Now,
consider a Borel subset $A\subset D^c$ in ${\bf R}^2$. For any sequence
of spins $(\tilde\sigma^N)$ in $\Omega_N$, 
$-\rho_N^\pm\le m_N^\pm(\tilde\sigma^N)\le \rho_N^\pm$,
and thus $\Psi_{B^N}(\tilde\sigma^N)\in A^c$ for $N$ large enough,
that is there exists $N_0(A)\in {\bf N}$ such that
$Q_N(A)=0$, $\forall N>N_0(A)$.

\section{Statistical approach}\label{s.statistical}
 \subsection{A Bayesian model}\label{s.bayesian}

In the heteropolymer case, the field $B^N=(b_i)$
is indexed by a word of length $N$ on the two letters alphabet
$\{A+T, G+C\}$. The DNA can be seen as a kind
of geometrical code where a word is glued on
the double helix. When the perimeter is fixed by setting
$V_N=U_{r,N}$,
${\cal C}_{r,N}=\{\sigma\in\Omega_N;\ \vert \sigma\vert=2r\}$ is a nonlinear code,
as a subset of $\Omega_N$, and tools from
information theory and bayesian statistics
are of great utility for computational issues.
We will see that the Hamiltonian
of the system models the a posteriori law
on the parameter space $\Omega_N$
given the observation $B^N$. First, consider
the {\it a priori } probability measure on the
parameter space given by
$$\nu_N(\sigma)=\frac{V_N(\sigma)\exp(-N\beta G(M_N(\sigma)))\prod_{i=1}^N
(\exp(-\sigma_i \beta b_{AT})+
\exp(-\sigma_i \beta b_{GC})}{Z_N(\nu_N)},$$
where $Z_N(\nu_N)$ denotes the related partition
function and
$V_N$ is an arbitrary probability measure on
$\Omega_N$.
Notice that $\nu_N$ takes into account the
level of superhelicity of the DNA through the
superhelical density $\kappa$, and that
$\nu_N$ corresponds to the Gibbs measure
associated with the  homopolymer with
field $\bar B$, given by
$$\bar b=\frac{b_{AT}+b_{GC}}{2},$$
that is
$$\nu_N(\sigma)=\frac{V_N(\sigma)\exp(-N\beta F_{\bar B}(\sigma))}{Z_N(\beta,\bar B,V_N)}.$$
The statistical model is as follows: a code word or a parameter
$\sigma^0$ is chosen at random with law $\nu_N$ on
${\rm supp}(V_N)$, and is sended through a noisy channel
with output alphabet $\{A+T,G+C\}$ and memoryless channel
statistics $P(\cdot\vert \sigma^0)$ given by
$$P(B^N\vert \sigma^0)=\prod_{i=1}^N p(b_i\vert\sigma_i^0)=\prod_{i=1}^N
\frac{\exp(-\sigma_i^0 \beta b_i)}{\exp(-\sigma_i^0 \beta b_{AT})+
\exp(-\sigma_i^0 \beta b_{GC})}.$$
The output distribution of $B^N$ is
$$q(B^N)=\sum_{\sigma\in \Omega_N}\nu_N(\sigma)P(B^N\vert \sigma),$$
and the {\it a posteriori distribution} on the parameter space is
the Gibbs distribution
$$\pi_{\beta,B^N,V_N}(\sigma)=\frac{V_N(\sigma)\exp(-N \beta
F_{B^N}(\sigma))}{Z_N(\beta,B^N,V_N)}.$$
Notice that 
$${\rm E}_q(\rho_N^+(B^N))=1-\theta +(2\theta-1)<M_N>_{\nu_N},$$
where $\theta=\exp(-\beta b_{AT})/(\exp(-\beta b_{AT})+\exp(-\beta b_{GC}))\ge 1/2$.
The law of $M_N(\sigma)$ under $\nu_N$
is subject to threshold phenomenon,
as shown in Section \ref{s.homo}. When
$V_N=U_{r_N,N}$ with $r_N/N\to 0$,
Theorem \ref{CriticalTransitions}
gives information on the limiting
support of $\nu_N$: $\kappa_c(\bar B)<\kappa<\bar\kappa_c(\bar B)$
implies that the law  of $M_N$ under $\nu_N$ converges
toward the point mass $\delta_{y_*(\bar B,\kappa)}$
with $y_*(\bar B,\kappa)\in (0,1)$, and the average
proportion of A+T bonds in $B^N$ under $q$
is given by $1-\theta +(2\theta-1)y_*(\bar B,\kappa)\in [1-\theta,\theta]$.

Having observed some word $B^N$, consider the Bayes estimator
for $\sigma^0$ under quadratic loss
$L(\sigma,\sigma')=\vert\vert \sigma-\sigma'\vert\vert^2$.
 Bayesian Theory 
 (see e.g. Robert (1992)) gives that
the optimal Bayes estimator under quadratic loss is
$$\hat\sigma_i=\frac{\sum_{\sigma\in {\cal C}_{r,N}}\sigma_i \nu_N(\sigma)
P(B^N\vert\sigma)}{\sum_{\sigma\in {\cal C}_{r,N}}\nu_N(\sigma)P(B^N\vert\sigma)},$$
and the estimated magnetization
$\sum_i \hat\sigma_i/N$ corresponds to our order parameter
$<m_N>_{\pi_{\beta,B^N,V_N}}$, showing that
Benham's model has an interesting statistical content.

 In the next Section, we imbeed Benham's model
in a bayesian segmentation model of current use
in bioinformatics, and provide new algorithms
with priors adapted to supercoiled DNA.

\subsection{Bayesian segmentation for strand separation}\label{s.segmentation}
\subsubsection{A two coins example}\label{Two}
Liu and Lawrence(1999) present a Bayesian model for
segmentation of biopolymers, and provide algorithms
for drawing samples from the various posterior laws
of the model, which might be of great interest
in the strand separation problem.  We start with
their {\it two types of coins} example, to fix ideas.
Suppose you know that in a coin tossing game of
length $N$, the first $A$ Bernoulli have a probability
of success $\theta_1$ and the next $N-A$ tosses have
a probability $\theta_2\ne\theta_1$ of getting
a head.  Let $y_{obs}$ be the observed sequence,
whith $h_1$ (resp. $h_2$) heads and $t_1$ (resp. $t_2$)
tails in the first (resp. second) part of the sequence.
The change point $A$ is treated as a missing data,
and has some prior law $g(a)$. The likelihood
of the observed data is given by
$$L(\theta_1,\theta_2;y_{obs},A=a)=\theta_1^{h_1}(1-\theta_1)^{t_1}
\theta_2^{h_2}(1-\theta_2)^{t_2}g(a).$$
In their work, the prior $\pi(\theta)$ for $\theta=(\theta_1,\theta_2)$
is a product measure associated with two independent
Beta random variables $B(\theta_1;\alpha_1,\beta_1)$
and $B(\theta_2;\alpha_2,\beta_2)$. Let
$p(\theta_1,\theta_2,a,y_{obs})$ be the joint law
of all variables, with
$$p(\theta_1,\theta_2,a,y_{obs})=L(\theta_1,\theta_2;y_{obs},A=a)\pi(\theta)g(a),$$
and
$$P(A=a,y_{obs})=\int\int p(\theta_1,\theta_2,a,y_{obs}){\rm d}\theta_1 {\rm d}\theta_2.$$
The following algorithm will converge and give samples from the
posterior law of $(\theta_1,\theta_2,A)$:
\begin{itemize}\label{Algo}
\item{}Fix $A=a$ and $\theta_2$, and draw $\theta_1$ from
its conditional posterior law $P(\theta_1\vert \theta_2, A=a,y_{obs})$,
to get the new $\theta_1$,
\item{}proceed similarly for $\theta_2$ to get the new $\theta_2$,
\item{}draw $A$ from its conditional posterior law
     $P(A=a\vert\ \theta,y_{obs})$ proportional to
     $\prod_{i=1,2}\theta_i^{h_i(a)}(1-\theta_i)^{t_i(a)}g(a)$,
     where $h_i(a)$ and $t_i(a)$ are the number of heads and tails
     contained in the $i$-th part of the sequence, $i=1,2$.
\end{itemize}
Coming back to the setting of Section \ref{s.bayesian}, choose
$V_N$ to be $U_{1,N}$, fixing the perimeter to $2r=2$. Benham's
model deals with a circular DNA, and they are
$N$ configurations of length $a$, and two change points.
Forgetting for a while this slight difference, we see
that the Bayesian model of Liu and Lawrence can be adapted
for Benham's model: $\sigma$ is the missing data or segmentation parameter $A$, and
the prior law $g(a)$ is just the a priori measure 
$\nu_N$ of Section \ref{s.bayesian}. Suppose that
$\sigma_i=+1$, $1\le i\le A$ and $\sigma_i=-1$,
$A<i\le N$. Set $B^N=y_{obs}$ and
fix the parameter $\theta$ to
\begin{equation}\label{Bare}
\bar\theta_1=\frac{\exp(-\beta b_{AT})}{\exp(-\beta b_{AT})+\exp(-\beta b_{GC})},\
\bar\theta_2=\frac{\exp(\beta b_{AT})}{\exp(\beta b_{AT})+\exp(\beta b_{GC})}=1-\bar\theta_1.
\end{equation}
The model of Section \ref{s.bayesian}
can be seen as a particular case of the segmentation model, and  the a posteriori law
of the change point $A$ is the Gibbs measure $\pi_{\beta,B^N,V_N}$.
For the deterministic model with constant $\bar\theta$, the last step
of the algorithm corresponds to sampling with
$\pi_{\beta,B^N,V_N}$.
Fye and Benham(1999) give algorithms for strand separation using
transfer matrices from statistical mechanics; the method can
be applied, thanks to the quadratic form appearing in the
prior $\nu_N$ (the transfer matrices have complex entries,
a consequence of the gaussian transform). The perimeter is not limited,
 and is penalized by the exponential weight
appearing in the one dimensional Ising Boltzman weight
$\exp(-4r\beta)$ for a perimeter of $2r$
 (notice that this way of penalizing too large perimeters
 was implemented in a recent work of Ramensky {\it et altri}(2000,2001)).
 Their method is applicable to situations where
the free energies for strand separation $b_{AT}$ and
$b_{GC}$ fluctuate: in real situations, chemical
reactions can alter these free energies.
In the randomized case, the Bayesian segmentation
model can also provide an interesting alternative  for studying the strand
separation problem with random free energies $b_{AT}$ and $b_{GC}$.

\subsubsection{The general case}
We adapt the setting of  Liu and Lawrence(1999)
and Ramensky {\it et altri}(2000,2001), and indicate
the main features of the model. The number
of domains or segments of the circular DNA
can be limited to $r_{\max}$. The missing data
is the spin $\sigma$,
with prior law $\nu_N$, given by $V_N$ and the homopolymer
of field $\bar b=(b_{AT}+b_{GC})/2$,
with free energies $b_{AT}$ 
 and
$b_{GC}$ given by formula (\ref{FreeEnergy}).
$\sigma$ can be seen as a juxtaposition
of droplets of $\pm$ spins arranged
around the discrete circle of length
$N$.

Suppose they are $2r$ domains
with $r$ positive droplets,
that is containing sites $i$
with
$\sigma_i=+1$, and $r$ negative
droplets. The parameter $\theta$
is here defined given $\sigma$.
We associate to every positive droplet
$\Lambda^+_{i_0}$ with
$\sigma_{i_0-1}=-1$, $\sigma_i=+1$,
$i_0\le i\le i_k$, and $\sigma_{i_k+1}=-1$
a family of $k+1$ i.i.d. Bernoulli $\varepsilon _i$,with
 $$P(\varepsilon_i=A+T)=\theta^{i_0,i_k}
 \hbox{ and }
 P(\varepsilon_i=G+C)=1-\theta^{i_0,i_k},$$
 of random
parameter $\theta^{i_0,i_k}$, which is chosen
according to some prior law $f^+$. Do the
same for negative droplets
for a random parameter $\theta^{j_0,j_l}$
of prior $f^-$,
with
 $$P(\varepsilon_j=A+T)=1-\theta^{j_0,j_l}
 \hbox{ and }
 P(\varepsilon_j=G+C)=\theta^{j_0,j_l}.$$
 The requirements might be
$${\rm E}_{f^+}(\theta^{i_0,i_k})={\rm E}_{f^-}(\theta^{j_0,j_l})=\bar\theta_1,$$
where the constants $b_{AT}$ and $b_{GC}$ appearing in
(\ref{FreeEnergy}) and (\ref{Bare}) can be taken as average values.
Positive droplets force the sample toward $A+T$
outcomes and conversely for negative droplets.
The priors $f^{\pm}$ can be chosen as in Liu and Lawrence(1999)
as Beta laws with the required expectations.

A second way of randomizing the parameters
consists in taking, independently for each segment,
two  positive random variables $b_{AT}(\omega)$
and $b_{GC}(\omega)$, distributed according to some law,
which might be motivated from thermodynamics or
biochemistry, with average values given by
 formula (\ref{FreeEnergy}).
 Draw $2r$ i.i.d. realizations of
these two random variables, and set, for each
segment,
$$\theta(\omega)=\frac{\exp(-\beta b_{AT}(\omega))}{\exp(-\beta b_{AT}(\omega))+\exp(-\beta b_{GC}(\omega)).}$$
The probability to get $A+T$ is $\theta(\omega)$ when the droplet is positive
and $1-\theta(\omega)$ when the droplet is negative.

Let $\pi(\theta\vert\sigma)$ be the prior law
given by the above construction, with
joint law
$\pi(\theta,\sigma)=\pi(\theta\vert\sigma)\nu_N(\sigma)$. Then the law of $B^N$
is
\begin{eqnarray*}
P(B^N)&=&\sum_{\sigma\in\Omega_N}\nu_N(\sigma)P(B^N\vert\sigma)\\
&=&\sum_{\sigma\in\Omega_N}\nu_N(\sigma)\int P(B^N\vert\theta,\sigma){\rm d}\pi(\theta\vert\sigma),
\end{eqnarray*}
where $P(B^N\vert \theta,\sigma)$ is the product
measure associated with the Bernoulli. The posterior
law of interest in the strand separation problem is just
$P(\sigma\vert B^N)=P(B^N\vert\sigma)/P(B^N)$. This last law is
define on the cube $\Omega_N$, of size $2^N$. The
observables of interest are the number of 
denatured bonds $M_N(\sigma)$ given by $m_N(\sigma)=2M_N(\sigma)-1$, and
the restricted magnetization
$m_N^+$ and
 $m_N^-$.
Information on the localization of denaturation can be obtained by
considering
 the proportion of broken A+T and G+C bonds
$$M_N^{\pm}=\frac{m_N^{\pm}+\rho_N^{\pm}}{2},$$
(see Section \ref{Hetero}).
Bayesian segmentation algorithms
like backward sampling, as given
in Liu and Lawrence(1999) or Schmidler {\it et altri}(2000),
can thus be applied to the strand separation problem
to study  local denaturation
as function of the various parameters,
taking into account the random fluctuations of
the free energies needed to denature A+T and G+C bonds.






\section*{References}

\noindent
R.~Baxter.
\newblock {\em Exactly Solved Models in Statistical Mechanics}
\newblock Academic Press, 1982.

\noindent
C.~Benham(1979)
\newblock Torsional stress and local denaturation in supercoiled DNA
\newblock {\em Proc. Natl. Acad. Sci. USA}, 76(8):3870-3874


\noindent
C.~Benham(1989)
\newblock Mechanics and Equilibria of Superhelical DNA.
\newblock In {\em Mathematical Methods for DNA Sequences}
\newblock Ed. M. Waterman
\newblock CRC Press.

\noindent
C.~Benham(1990)
\newblock Theoretical analysis of heteropolymeric transitions in superhelical 
DNA molecules of specified sequences.
\newblock {\em J. Chem. Phys.}, 92(10):6294--6305

\noindent
C.~Benham(1992)
\newblock Energetics of the Strand Separation Transition in Superhelical DNA
\newblock {\em J. Mol. Bio.}, 225:835-847

\noindent
C.~Benham(1996)
\newblock Theoretical analysis of the helix-coil transition in positively 
superhelical DNA at high temperature
\newblock {\em Phys. Rev. E},53(3):2984--2987.

\noindent
P.~Clote and R.~Backofen
\newblock {\em Computational Molecular Biology: An Introduction }
\newblock  Wiley, 2000.

\noindent A.~Dembo and O.~Zeitouni
\newblock {\em Large Deviations Techniques and Applications}
\newblock Jones and Bartlett, 1992.

\noindent J.~Deuschel and D.~Stroock
\newblock {\em Large Deviations}
\newblock Academic Press. 1989.

\noindent R.~Ellis
\newblock {\em Entropy, Large Deviations and Statistical Mechanics}
\newblock Springer 1985.

\noindent W.~Feller
\newblock {\em An Introduction to Probability Theory and Its Applications, Vol.
I}
\newblock Wiley, 1971.

\noindent
R.~Fye and C.~Benham(1999)
\newblock Exact method for numerically analyzing a model of local
denaturation in superhelically stressed DNA
\newblock    {\em Phys. Rev. E}, 59:3408.

\noindent B.~Lewis
\newblock {\em Genes V}
\newblock Oxford University Press. 1994.

\noindent J.S.~Liu and C.E.~Lawrence(1999)
\newblock Bayesian inference on biopolymer models
\newblock {\em Bioinformatics} 15(1):38-52

\noindent J.S.~Liu, A.F.~Neuwald and C.E.~Lawrence
\newblock Markovian structures in biological sequence alignments
\newblock Preprint

\noindent P.~Mathieu and P.~Picco(1998). Metastability and convergence
to equilibrium for the random field Curie Weiss model.
{\em J. Stat. Phys.} 91(3/4): 679-732.

\noindent V.E.~Ramensky, V.J.~Makeev, M.A.~Roytberg and V.G.~Tumanyan(2000)
\newblock DNA segmentation through the Bayesian approach
\newblock {\em Journal of Computational Biology}. 7(1/2): 215-231

\noindent V.E.~Ramensky, V.J.~Makeev, M.A.~Roytberg and V.G.~Tumanyan(2001)
\newblock Segmentation of long genomic sequences into domains with
homogeneous composition with BASIO software
\newblock {\em Bioinformatics} 17(11):1065-1066.

\noindent C.~Robert
\newblock {\em The Bayesian Choice}
\newblock Springer 1992.

\noindent R.T.~Rockafellar
\newblock {\em Convex Analysis}
\newblock Princeton University Press. 1972

\noindent
H.~Sun, M.~Mezei, R.~Fye and C.~Benham(1995)
\newblock Monte-Carlo analysis of conformational transitions in superhelical DNA
\newblock    {\em J. Chem. Phys}, 103(19):8653--8665.

\noindent S.C.~Schmidler, J.S.~Liu and D.L.~Brutlag(2000)
\newblock Bayesian segmentation of protein secondary structure
\newblock {\em Journal of Computational Biology}. 7(1/2): 233-248



\end{document}